# Empirical LCAO parameters for π molecular orbitals in planar organic molecules

L. Hawke[1], G. Kalosakas[1], C. Simserides[2]

[1]Materials Science Department, University of Patras, GR-26504, Rio, Greece

[2] Institute of Materials Science, NCSR Demokritos, GR-15310, Athens, Greece

**Abstract**

We present a parametrization within a simplified LCAO model (a type of Hückel model) for the description of π molecular orbitals in organic molecules containing π-bonds between carbon, nitrogen, or oxygen atoms with sp$^2$ hybridization, which we show to be quite accurate in predicting the energy of the highest occupied π orbital and the first π-π* transition energy for a large set of organic compounds. We provide four empirical parameter values for the diagonal matrix elements of the LCAO description, corresponding to atoms of carbon, nitrogen with one $p_z$ electron, nitrogen with two $p_z$ electrons, and oxygen. The bond-length dependent formula (proportional to $1/d^2$) of Harrison is used for the non-diagonal matrix elements between neighboring atoms. The predictions of our calculations have been tested against available experimental results in more than sixty organic molecules, including benzene and its derivatives, polyacenes, aromatic hydrocarbons of various geometries, polyenes, ketones, aldehydes, azabenzenes, nucleic acid bases and others. The comparison is rather successful, taking into account the small number of parameters and the simplicity of the LCAO method, involving only $p_z$ atomic orbitals, which leads even to analytical calculations in some cases.

## 1. Introduction

Theoretical and experimental efforts for the determination of the electronic structure of organic molecules started as soon as quantum mechanics was established as the fundamental theory for the microscopic description of matter. These efforts, except for the evaluation of the energy eigenvalues of the electronic states, were concerned also with other aspects, like for example the determination of the symmetry of each electronic state, the assignment of electronic transitions (e.g. singlet-singlet or singlet-triplet transitions, Rydberg transitions, π-π* transitions etc.), and the calculation of the oscillator strength of the transitions. Apart from basic knowledge and the numerous applications of planar organic molecules containing atoms with sp$^2$ hybridization, the π molecular electronic structure of such compounds is involved in several biological functions. For example, we mention chlorophyll in photosynthesis, the retinal molecule involved in vision or in photon-driven ion pumps like bacteriorhodopsin [1], and many molecules with photobiological functions such as vitamin A, vitamin D precursors, carotene etc., containing polyene chromophores [2]. Also new organic semiconductors based in pentacene and other hydrocarbon molecules have attached an enormous interest regarding their use in molecular electronics [3].

Experimental investigations of the electronic structure of organic molecules started very early, by performing absorption measurements. The ultraviolet absorption spectra of eighteen pyridines and purines [4], fourteen ethylenic hydrocarbon molecules [5], and 1,3 cyclohexadiene [6] have been measured already in 30s. Later Platt and Klevens presented the spectra of several alkylbenzenes [7], some ethylenes and acetylenes [8], and seventeen polycyclic aromatic hydrocarbons composed of fused benzene rings [9] (e.g. phenanthrene and chrysene). At the same period spectra were taken from naphthalene and biphenyl derivatives [10], m- and o-disubstituted benzene derivatives [11], and mono-substituted and p-disubstituted benzene derivatives [12]. Absorption measurements continued with the same intensity the following decades [13-15]. During the last thirty years, several new methods emerged for the measurement of the electronic structure of molecules. Some of them are experimentally easier from conventional absorption spectroscopy and may be able to probe optically forbidden transitions. For instance singlet-triplet transitions can be easily assigned by such methods. Particularly, the electron impact method has been applied in 1,3,5 hexatriene [16], resonant enhanced multiphoton ionization in pyrrole, N-methyl pyrrole, and furan [17], electron scattering spectroscopy in propene [18] and isobutene [19], and cavity ring-down spectroscopy in 1,3 butadiene [20].

Early theoretical efforts to describe π molecular structure have been done by Hückel in 30s [21] (for a recent review on Hückel theory and its aspects see Ref. [22] and also references therein for important contributions). Platt predicted the first two electronic transitions of sixteen conjugated molecules by using the LCAO method [23] and also tried to summarize and justify general laws that govern electronic spectra [24]. Another theoretical attempt at the same period was done by Pariser and Parr who predicted the first main visible or ultraviolet absorption bands of benzene and ethylene [25], butadiene, pyridine, pyrimidine, pyrazine, and s-triazine [26], and several polyacenes [27]. Their semi-empirical theory, known as PPP theory, was based on antisymmetrized products of molecular orbitals, obtained using the LCAO approximation. Other models came out in the following decades, like for example the CNDO/S2 spectroscopic model that was applied for the description of the electronic excitation spectra of polyacenes (naphthalene to pentacene), providing results in good agreement with experimental data [28]. In the last two decades the theoretical efforts were focused on more accurate calculations from first principles. Such methods have been applied in many organic molecules, like for example in benzene [29, 30], azabenzenes [29], heptacene [30], naphthalene, anthracene, tetracene and hexacene [30, 31], pentacene [30, 31, 32], pyrrole [33, 34], furan [34], butadiene and hexatriene [35], and cyclic ketones and thioketones [36].

Although methods from first principles some times -depending on the used basis set or the method itself- are not so accurate for all orbitals (especially for the unoccupied ones), in general they can provide very successful predictions of the electronic structure. Therefore, methods from first principles are of extreme importance for the interpretation of molecular electronic spectra. However, as also happens in the experimental observations, a computationally demanding first principles calculation may not give particular insights at the underlying mechanism responsible for the obtained result. On the contrary, much simpler semi-empirical methods, usually containing a few parameters, even though less accurate, may be in the position to provide a more fundamental understanding of the electronic structure and its dependence on the physical properties of the system. Excellent demonstrations of these ideas are provided by the impressive work of Harrison [37], who was able to account for various properties (ranging from dielectric, to conducting, elastic etc.) of different categories of solids using such a simple approach, and by Streitwieser [38], who summarized early efforts along these lines regarding molecular properties. Inspired by these works, the aim of this article is the evaluation of the electronic structure of π molecular orbitals, by using the linear combination of atomic orbitals (LCAO) method including only $p_z$-type orbitals (like in Hückel theory) and a minimal unified set of parameters for describing a relatively large number of planar organic molecules.

## 2. Theoretical method

Atoms in planar organic molecules with sp$^2$ hybridization have their $p_z$ atomic orbitals perpendicular to the molecular plane. The electrons that occupy these orbitals are eventually delocalized. The LCAO method provides a very simple way to calculate π molecular orbitals, which approximately describe these delocalized electrons. In this approximation the corresponding molecular wavefunction $\psi(\vec{r})$ is a linear combination of the $p_z$ atomic orbitals from each atom, or, in the context used in the present work, of atomic-like orbitals $p$ which resemble the $p_z$ atomic orbitals:

$$\psi(\vec{r}) = \sum_{i=1}^{N} c_i p_i(\vec{r}) \ . \tag{1}$$

The summation index, $i$, runs among the $N$ atoms of the molecule, which contribute $p_z$ electrons in π bonds. Here we ignore all other orbitals (including the sp$^2$ hybrids) and consider only the Hamiltonian in the subspace of $p$ orbitals.

Multiplying the Schrödinger equation,

$$\hat{H}\psi = E\psi \ , \tag{2}$$

with the conjugate atomic-like orbital $p_j^*(\vec{r})$ and integrating, we obtain the linear system

$$\sum_{i=1}^{N} \left[(H_{ji} - E\delta_{ji})c_i\right] = 0, \ \text{for} \ j = 1, 2, \ldots, N \ . \tag{3}$$

Here we have assumed orthogonality of the $p$ orbitals located in different atoms, i.e., $\int p_j^*(\vec{r}) p_i(\vec{r}) \, d^3r = \delta_{ji}$, where $\delta$ is the delta of Kronecker, otherwise the corresponding overlap integral should be included in Eq. (3). The Hamiltonian matrix elements $H_{ji}$ are given by

$$H_{ji} = \int p_j^*(\vec{r}) \, \hat{H} \, p_i(\vec{r}) \, d^3r \ . \tag{4}$$

Thus, in this approximation we obtain the coefficients, $c_i$, which provide the π molecular orbitals through Eq. (1), and the corresponding energy eigenvalues $E$ by numerical diagonalization of the Hamiltonian matrix, as it can be seen from Eq. (3). The only information needed in this approach is the values of the matrix elements, $H_{ij}$.

Regarding the diagonal matrix elements, $H_{ii}$, depending on the atom in which the index $i$ is referred, we use the values $\varepsilon_C = -6.7$ eV for carbon, $\varepsilon_{N_2} = -7.9$ eV for nitrogen with one electron in the $p_z$ atomic orbital (i.e. with coordination number 2), $\varepsilon_{N_3} = -10.9$ eV for nitrogen with two electrons in the $p_z$ atomic orbital (i.e. with coordination number 3), and $\varepsilon_O = -11.8$ eV for oxygen. We arrived at these empirical values after a series of simulations of the electronic structure of various organic molecules. Initially we tried to use the ionization energies of the elements C (-11.26 eV), O (-13.62 eV), N (-14.53 eV), as it is usually chosen for the diagonal matrix elements. However, for all the molecules examined, using the ionization energies of the elements led to large deviations from the experimentally known molecular ionization energies of the highest occupied π orbital. Therefore, we used $\varepsilon_C$ as a free parameter and tried to fit many cyclic and non-cyclic hydrocarbons. It turned out that the value $\varepsilon_C = -6.7$ eV, results in good agreement (within less than 13% deviation) between the calculated and the experimental π ionization energies of the investigated molecules. This value almost coincides with the analytically derived value of $-6.8$ eV, which provides the exact experimental value of the benzene ionization energy. Next, we fixed $\varepsilon_C = -6.7$ eV and examined organic molecules containing nitrogen and oxygen atoms, with emphasis in the nucleic acids' bases. It turned out that the best choice was the above mentioned values of $\varepsilon_{N_2}$, $\varepsilon_{N_3}$, and $\varepsilon_O$. At this point we mention that different LCAO diagonal energies for two types of nitrogen atoms, distinguished by their coordination numbers as $N_2$ and $N_3$, i.e. with one or two electrons in the $p_z$ atomic orbital, have also been used in the literature [38,39].

The nondiagonal (i.e. interatomic) matrix elements $H_{ij}$ (known also as resonance integrals) are zero if the indices $i$ and $j$ refer to atoms without a direct bond between them, while for neighboring bonded atoms we use the expression proposed by Harrison [37]:

$$H_{ij} = -0.63 \frac{\hbar^2}{m d_{ij}^2}, \text{ for } i, j \text{ referring to neighboring bonded atoms.} \quad (5)$$

Here, $m$ is the electron mass and $d_{ij}$ is the length of the bond between the atoms $i, j$. Harrison' formula is universal and applies to corresponding matrix elements between different elements. The proportionality of the matrix elements to $1/d^2$ is not valid for arbitrary distances, but only at distances near to the equilibrium interatomic distances in matter. We remark that this expression for $H_{ij}$ describes the matrix elements between adjacent $p_z$-type orbitals under the hypothesis that their overlap is ignored. This is consistent with our previously mentioned assumption in deriving Eq. (3). Harrison's interatomic matrix elements are very popular among physicists because, as we already mentioned, they can successfully describe a large variety of properties of materials within a simple LCAO approximation [37]. Such a dependence of the interatomic matrix elements (proportional to $1/d^2$) has not been used by chemists in the application of LCAO in molecules, where the rather more complicated Wolfsberg-Helmholz expression [40] is widely applied. However, since the interatomic distances are similar in molecules and solids, one expects that Harrison's matrix elements (5) can be also applied in molecules. Their advantage, compared to the well-known Wolfsberg-Helmholz interatomic matrix elements, is that they are considerably simpler and readily applied when the interatomic distance (bond length) $d$ is known. Further, as we show in this work, they can be rather successfully used for estimating the energy of the highest occupied $\pi$ orbital and the first $\pi$-$\pi^*$ transition in a large number of organic molecules. The geometries and the interatomic distances $d_{ij}$ in all the theoretically investigated molecules in this work are obtained from the NIST website [41]. We mention that our LCAO method with the above values of diagonal and nondiagonal matrix elements can be considered a type of Hückel model with explicit bond-length dependence of the resonance integrals.

The energy eigenvalues obtained from the numerical diagonalization of the Hamiltonian matrix correspond to the electronic spectrum of π molecular orbitals. Then the occupied and unoccupied π orbitals of the organic molecule can be found by counting all the $p_z$ electrons contributed by the atoms of the molecule and arrange them successively in couples of different spin in accordance to Pauli principle. Additionally the π-π* transitions can be obtained. In some molecules like benzene and polyacenes the HOMO-LUMO gap (i.e. the energy gap between the highest occupied molecular orbital and the lowest unoccupied molecular orbital) corresponds to a π-π* transition. However, this is not always the case, as we discuss later, for example in the case of polyenes or some heterocyclic organic compounds, where the HOMO-LUMO gap is not a π-π* transition.

## 3. Results and Discussion

*3.1 Benzene, polyacenes, and aromatic hydrocarbons of various geometries*

In Table I we present our results for benzene, polyacenes and a number of aromatic hydrocarbons with many rings and various architectures. In particular, we show the calculated ionization energy of the highest occupied π molecular orbital ($\pi$HOMO), $I\pi_{th}$, the corresponding experimental value, $I\pi_{exp}$, and the respective % deviation,

$$\text{error}_1 = \frac{|I\pi_{th} - I\pi_{exp}|}{I\pi_{exp}} \times 100, \qquad (6)$$

for each organic molecule of the table. Furthermore the calculated energy of the first excited π* orbital ($\pi$LUMO), $L\pi_{th}$, is displayed along with the resulting theoretical π-π* energy gap, π-π*$_{th}$, the experimental one, π-π*$_{exp}$, and the corresponding % deviation,

$$\text{error}_2 = \frac{|(\pi\text{-}\pi^*_{th}) - (\pi\text{-}\pi^*_{exp})|}{\pi\text{-}\pi^*_{exp}} \times 100 \qquad (7).$$

The experimental values in this and the following tables correspond to vertical ionizations or excitations (i.e. without a change in the structure of the molecule). It must be mentioned that the first π-π* transition in this kind of molecules of Table I corresponds to the HOMO-LUMO gap.

In respect to the ionization energy (the energy that must be given for the evacuation of an electron from the highest occupied π molecular orbital), the LCAO predicted results are in very good agreement with the experimental data. The biggest deviation (12.5%) is in hexacene and the smaller ones in benzene and naphthalene. For larger polyacenes (anthracene and tetracene) the relative error is between 5-7.5%, while it exceeds 10% for even larger polyacenes (pentacene and hexacene). In most aromatic hydrocarbons of Table

I with nonlinear architecture the deviation is larger than 8%, but less than 12% (apart from benzo[p]hexaphene-naphtho(2',3':1,2)pentacene, where it is 12.1%).

Regarding the π-π* energy gap, the deviations are larger and in certain cases are more than 40%. In particular in pentacene, hexacene and benzo[p]hexaphene-naphtho(2',3':1,2)pentacene the relative errors are 43-45%, 44% and 42%, respectively. In other six molecules the deviations are between 30-40%, while in the remaining fourteen from the organic compounds of Table I where experimental data are available, the relative error is below 30%. The lowest deviation for the π-π* energy gap is found in benzene. It must be mentioned here that the energy gap decreases for polyacenes as the number of benzene rings increases. This is due to a wider splitting of the energy states resulting from the $p_z$ atomic orbitals, as the number of atoms increases. This experimentally verified trend is captured from our theoretical calculations, even though the corresponding deviations increase for larger polyacenes.

A comparison of the calculated first π-π* transition between the used model and methods from first principles can be made for some molecules of Table I. Particularly for benzene, the LCAO predicts a HOMO-LUMO gap of 5.0 eV and several first principles methods 4.84 eV [42], 5.14 eV [29] (a state of the art *ab initio* calculation), and 5.24-5.28 eV [30]. For the polyacenes first principles calculations are more accurate. For example in naphthalene the LCAO gives a value of 3.2 eV and various methods from first principles predict 4.09-5.27 eV [31], 4.38-4.88 eV [43], and 4.27 eV [30], which are closer to the experimental value of 3.9-4.0 eV, except for the extreme values of 4.88 eV and 5.27 eV. Furthermore, a comparison between LCAO predictions, experimental data, and first principles calculations included in Refs. [30], [31], and [43] for anthracene, tetracene, pentacene and hexacene confirms this conclusion.

**Table I.** Benzene, polyacenes, and aromatic hydrocarbons of various geometries. The first column depicts the organic molecule. The second and the third column present the experimental and theoretical, respectively, ionization energy of the highest occupied π molecular orbital. The fourth column has the % relative error between the experimental and theoretical values. In the fifth column the evaluated energy of the first excited π* orbital is shown. The sixth and the seventh columns include the theoretical and the experimental π-π* energy gap, respectively. The last column presents the % deviation between the calculated and experimental π-π* gaps.

| Organic molecule | $I\pi_{exp}$ (eV) | $I\pi_{th}$ (eV) | error$_1$ % | $L\pi_{th}$ (eV) | π-π*$_{th}$ (eV) | π-π*$_{exp}$ (eV) | error$_2$ % |
|---|---|---|---|---|---|---|---|
| 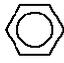 Benzene | 9.2-9.3 [44-55] | 9.2 | 0-1.1 | -4.2 | 5.0 | 4.7 [56]-4.9 [57] | 2-6 |
| 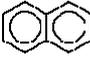 Naphthalene | 8.1-8.3 [46,49,50, 53,58-61] | 8.3 | 0-2.5 | -5.1 | 3.2 | 3.9 [56]-4.0 [57] | 18-20 |
| 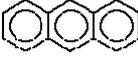 Anthracene | 7.4 [49,50,60,62,63] | 7.8 | 5.4 | -5.6 | 2.2 | 3.3 [56,57] | 33 |
| 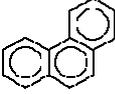 Phenanthrene | 7.9 [49,50,62,64-66] | 8.3 | 5.1 | -5.1 | 3.2 | 3.5 [9] | 9 |
| 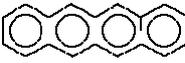 Naphthacene(or Tetracene) | 7.0 [49,67] | 7.5 | 7.1 | -5.9 | 1.6 | 2.6 [56,57] | 38 |

| Structure | | | | | | | |
|---|---|---|---|---|---|---|---|
| 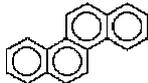 Chrysene | 7.6 [49,50,67] | 8.1 | 6.6 | -5.3 | 2.8 | 3.4 [56] | 18 |
| 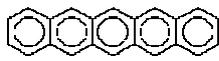 Pentacene | 6.6 [49] | 7.3 | 10.6 | -6.1 | 1.2 | 2.1 [57]-2.2 [56] | 43-45 |
| 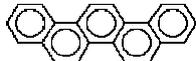 Picene | 7.5 [49,50] | 8.1 | 8.0 | -5.3 | 2.8 | 3.3 [68] | 15 |
| 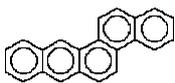 3,4-benzotetraphene | 7.2 [49] | 7.8 | 8.3 | -5.6 | 2.2 | 3.3 [68] | 33 |
| 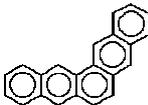 Pentaphene | 7.3 [49,50] | 7.9 | 8.2 | -5.5 | 2.4 | - | - |
| 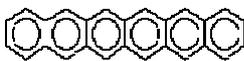 Hexacene | 6.4 [49] | 7.2 | 12.5 | -6.2 | 1.0 | 1.8 [69] | 44 |
| 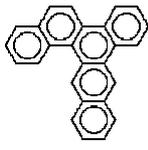 Naphtho[2,3-g] chrysene | 7.2 [49] | 7.9 | 9.7 | -5.5 | 2.4 | 3.1-3.3 [68] | 23-27 |
| 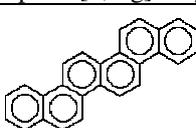 Benzo[c]picene-fulminene | 7.2 [50] | 8.0 | 11.1 | -5.4 | 2.6 | 3.2 [68] | 19 |
| 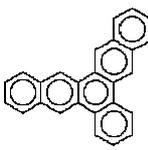 6,7benzopentaphene-Benzo[h]pentaphene | 7.4 [49] | 8.0 | 8.1 | -5.4 | 2.6 | 3.2 [68] | 19 |
| 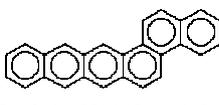 Naphtho[2,1a]naphthacene-naphtho(2',1':1,2)tetracene | 6.8 [68] | 7.6 | 11.8 | -5.9 | 1.7 | 2.7 [68] | 37 |
| 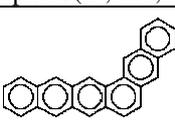 Hexaphene | 6.9 [68] | 7.6 | 10.1 | -5.8 | 1.8 | 2.7 [68] | 33 |
| 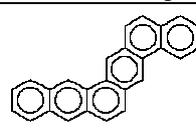 3,4benzopentaphene-Benzo[c]pentaphene | 7.2 [68] | 7.9 | 9.7 | -5.5 | 2.4 | 3.0 [68] | 20 |

| Structure | | | | | | | |
|---|---|---|---|---|---|---|---|
| 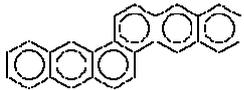 Dibenzo[b,k]chrysene-anthraceno(2',1':l,2)anthracene | 7.0 [68] | 7.7 | 10.0 | -5.7 | 2.0 | 3.0 [68] | 33 |
| 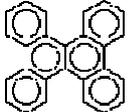 Dibenzo[g,p]chrysene | 7.2 [49] | 8.0 | 11.1 | -5.4 | 2.6 | - | - |
| 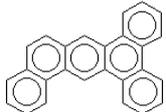 Tribenz[a,c,h]Anthracene | 7.4 [49,50] | 8.1 | 9.5 | -5.3 | 2.8 | - | - |
| 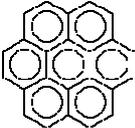 Coronene | 7.3 [50,70] | 8.1 | 11.0 | -5.3 | 2.8 | 3.0 [56] | 7 |
| 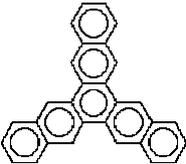 Trinaphthylene-naphtho (2',3':6,7)Pentaphene | 7.4 [68] | 8.1 | 9.5 | -5.3 | 2.8 | 3.1 [68] | 10 |
| 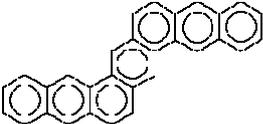 Naphtho[2,3c]Pentaphene-Naphtha(2',3':3,4)pentaphene | 7.0 [68] | 7.8 | 11.4 | -5.6 | 2.2 | 2.9 [68] | 24 |
| 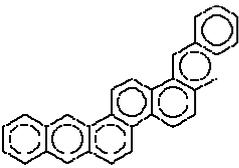 Dibenzo[b,n]picene-2,3:8,9-dibenzopicene | 7.2 [68] | 7.8 | 8.3 | -5.6 | 2.2 | 2.9 [68] | 24 |
| 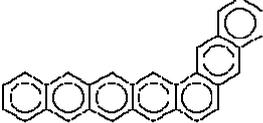 Benzo[p]hexaphene-naphtho (2',3':1,2)pentacene | 6.6 [68] | 7.4 | 12.1 | -6.0 | 1.4 | 2.4 [68] | 42 |
| 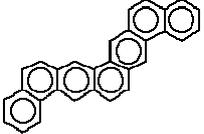 Dibenzo[c,m]pentaphene | 7.1 [49,50] | 7.8 | 9.9 | -5.6 | 2.2 | - | - |
| 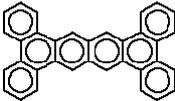 Tetrabenzo[a,c,j,l]naphthacene | **7.0** [50] | **7.8** | **11.4** | **-5.6** | **2.2** | - | - |
| 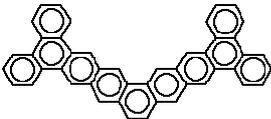 Tetrabenzo[a,c,g,s]heptaphene | **6.9** [49] | **7.7** | **11.6** | **-5.7** | **2.0** | **2.6** [68] | **23** |

*3.2 Polyenes*

The results for several polyenes are shown in Table II. In most of these molecules the π molecular orbitals can be obtained readily analytically in our approach, since there are only a couple of sp$^2$ hybridized atoms forming π-bonds. Here, the maximum deviation between the predicted and experimental πHOMO ionization energy is around 11-12%. The highest deviations 11.4-12% and 9.5-11.2% are found in 2-butene,2,3-dimethyl and ethylene, respectively. These are the only deviations among the investigated molecules exceeding 10%, while the others are no more than 8%.

The theoretical π-π* energy gap exhibits larger deviations from the experimental value, as happened in the previous subsection. The lowest deviation is around 19% in 1-propene,2-methyl (isobutene) and the highest 42%-43% in 1,3,5-hexatriene. Here these higher deviations can be explained from the existence of Rydberg states, which interpolate energetically in between the π states. It must be mentioned that the actual HOMO-LUMO gap in these molecules is not a π-π* transition, but a transition from a π state to a Rydberg state. For a better prediction of the electronic spectra of these molecules within the LCAO approximation, the higher-energy atomic states can be included in the atomic orbital expansion of the molecular wavefunction. In the expansion used here, for simplicity, the π molecular orbitals were considered as isolated (far energetically) in respect to the other orbitals and only the $p_z$ atomic states were included in the LCAO method. It must be also mentioned that for ethylene there is a dispute on whether the π-π*$_{exp}$ value shown in Table II corresponds to a vertical transition or to a twisted configuration of the molecule. The latter hypothesis is supported by a number of theoretical studies of increasing accuracy, which have led to a final estimate of about 8.0 eV for the vertical transition energy [71,72,73].

The exclusion of Rydberg states from our consideration explains also why first principles methods may provide a better agreement with the observed values when applied in ethylene, 1,3-butadiene, and 1,3,5-hexatriene. Especially for ethylene such methods estimate the first π-π* transitions in the region of 7.97-8.54 eV [74], which is closer to the experimental value than the present LCAO estimation of 5.5 eV. In 1,3-butadiene the experimental value is 5.9 eV and first principles calculations predict 6.12-8.54 eV [74]. By using the LCAO, the prediction is 3.6 eV which is less accurate than 6.12 eV but no worse than 8.54 eV. For 1,3,5-hexatriene first principles evaluations range between 5.01-7.36 eV [74]. Again the upper limit of this region is no more accurate than the simple LCAO calculation (2.8 eV).

**Table II.** Polyenes. The columns represent the same quantities as in Table I.

| Organic molecule | $I\pi_{exp}$ (eV) | $I\pi_{th}$ (eV) | error$_1$ % | $L\pi_{th}$ (eV) | $\pi$-$\pi^*_{th}$ (eV) | $\pi$-$\pi^*_{exp}$ (eV) | error$_2$ % |
|---|---|---|---|---|---|---|---|
| 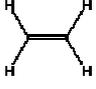 Ethylene | 10.5-10.7 [47, 75-78] | 9.5 | 9.5-11.2 | -4.0 | 5.5 | 7.6 [75]-7.7 [79] | 28-29 |
| 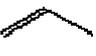 Propene | 9.7-10.2 [47,75-77, 80,81] | 9.4 | 3.1-7.8 | -4.0 | 5.4 | 7.2 [82] | 25 |
| 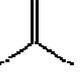 1-propene,2-methyl | 9.4-9.5 [80,83] | 9.4 | 0-1.1 | -4.0 | 5.4 | 6.7 [84] | 19 |
| 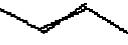 2-butene | 9.1 [45] | 9.4 | 3.3 | -4.0 | 5.4 | - | - |
| 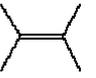 2-butene,2,3-dimethyl | 8.3-10.5 [45,85] | 9.3 | 11.4-12.0 | -4.1 | 5.2 | - | - |
| 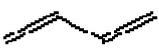 1,3-butadiene | 9.0 [86] –9.1 [75] | 8.5 | 5.6-6.6 | -4.9 | 3.6 | 5.9 [87,88] | 39 |
| 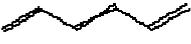 1,3,5-hexatriene | 8.3 [75] | 8.1 | 2.4 | -5.3 | 2.8 | 4.8 [56]-4.9 [16,89] | 42-43 |

*3.3 Benzene derivatives and azulene*

In the Table III below we report our calculations for some benzene derivatives. Once more, the evaluated ionization energies of the highest occupied π molecular orbital are pretty close to the corresponding experimental values. The highest relative errors appear in tetralin and p-xylene (8.2-9.5% and 7.0-9.5% respectively), while in most of the remaining cases the relative errors are below 5%.

Regarding the first π-π* transition, little information could be found. Particularly we were able to find results only for toluene, styrene, p-xylene and azulene. In toluene the prediction is close to the experimental value (the relative error is 6%). In p-xylene and styrene the deviation between theoretical and experimental data is 11% and 18%, respectively, while in azulene the relative error is around 28%.

**Table III.** Benzene derivatives and azulene. The columns show the same quantities as in Table I.

| Organic molecule | $I\pi_{exp}$ (eV) | $I\pi_{th}$ (eV) | error$_1$ % | $L\pi_{th}$ (eV) | $\pi$-$\pi$*$_{th}$ (eV) | $\pi$-$\pi$*$_{exp}$ (eV) | error$_2$ % |
|---|---|---|---|---|---|---|---|
| 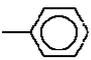 Toluene | 8.8-9.0 [44,47,48,51,52,90] | 9.2 | 2.2-4.5 | -4.2 | 5.0 | 4.7 [91] | 6 |
| 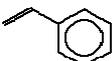 Styrene | 8.5-8.6 [44,53,61,92] | 8.5 | 0-1.2 | -4.9 | 3.6 | 4.4 [91,93] | 18 |
| 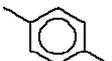 p-xylene | 8.4-8.6 [44,52,94] | 9.2 | 7.0-9.5 | -4.2 | 5.0 | 4.5 [91] | 11 |
| 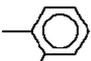 Benzene,1,2-dimethyl | 8.6-8.8 [44,52,59,95] | 9.2 | 4.5-7.0 | -4.2 | 5.0 | - | - |
| 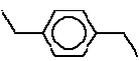 Benzene,1,4-diethyl | 8.4 [96] | 8.2 | 2.4 | -5.2 | 3.0 | - | - |
| 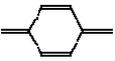 1,4-cyclohexadiene, 3,6-bis(methylene)- | 7.9 [97] | 7.7 | 2.5 | -5.7 | 2.0 | - | - |
| 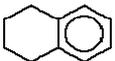 Tetralin | 8.4 [95]-8.5 [98] | 9.2 | 8.2-9.5 | -4.2 | 5.0 | - | - |
| 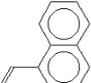 1-ethenylnaphthalene | 7.9-8.0 [99] | 8.1 | 1.3-2.5 | -5.3 | 2.8 | - | - |
| 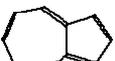 Azulene | 7.4 [59,100] | 7.9 | 6.8 | -5.6 | 2.3 | 1.8 [101] | 28 |
| 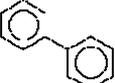 Biphenyl | 8.3 [65]- 8.4 [64] | 8.5 | 1.2-2.4 | -4.9 | 3.6 | - | - |
| 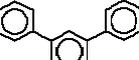 m-terphenyl | 8.1 [102] | 8.4 | 3.7 | -5.0 | 3.4 | - | - |

### 3.4 Organic compounds containing nitrogen atoms

Table IV presents the same quantities with the previous tables for cyclic heteroatomic organic compounds containing nitrogen atoms. The results for the πHOMO ionization energies show a relatively good agreement with the experimental values. In particular, only in three among the seven molecules of this table the deviation is larger than 10%. The biggest deviation is 17.5% in pyrimidine and the next one is 14.5% in 1,3,5triazine (or s-triazine). It must be mentioned here that in some of the molecules of Table IV the HOMO is not a π molecular orbital, but an antibonding n orbital, originating from the nitrogen atoms contained in these molecules. Generally, the highest deviations appear in those molecules, apparently because of the interaction between π and n orbitals. Many efforts have been done in order to clarify whether the HOMO of these molecules is a n or a π state. It seems that the HOMO orbital is ordered as a n state in 1,3,5 triazine [103], pyrimidine, and pyridazine [54].

Regarding the π-π* energy gaps, small deviations are obtained using the simple LCAO method in this family of molecules. For the azabenzenes (1,3,5triazine, pyrimidine, pyridazine, pyridine) the deviations do not exceed 6%. In azabenzenes also the results from first principles calculation are similar to the experimental ones. Such first principles predictions are 5.33-5.80 eV for 1,3,5triazine, 4.93-5.44 eV for pyrimidine, 4.86-5.31 eV for pyridazine, and 4.84-5.22 eV for pyridine [29]. We see that for azabenzenes, the LCAO method, although simple, is almost as accurate as methods from first principles for the prediction of the first π-π* transition. Further, we mention that in the case of 1,3,5triazine - because of its high symmetry- the LCAO π molecular electronic structure can be readily obtained analytically, since the original $6 \times 6$ matrix that has to be diagonalized ends up to a $2 \times 2$ matrix by virtue of the Bloch theorem.

For pyrrole, we remark that despite the fact that various types of experiments and theoretical investigations have been devoted to its study, a detailed assignment of many transitions has not been achieved, yet. Earlier experiments indicated that the first π-π* transition is at 5.22 eV [15], but later calculations proposed that the specific transition is of different nature [33b], or others defined the first π-π* transition in a higher energy region [33a, 34]. Information about the π-π* transitions of 1H-imidazole and 1H-imidazole,2-methyl could not be found.

**Table IV.** Organic compounds containing nitrogen atoms. The columns represent the same quantities as in Table I.

| Organic molecule | $I\pi_{exp}$ (eV) | $I\pi_{th}$ (eV) | error$_1$ % | $L\pi_{th}$ (eV) | π-π*$_{th}$ (eV) | π-π*$_{exp}$ (eV) | error$_2$ % |
|---|---|---|---|---|---|---|---|
| 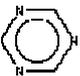 1,3,5-triazine | 11.7 [103] | 10.0 | 14.5 | -4.6 | 5.4 | 5.6 [56]-5.7 [104] | 4-5 |
| 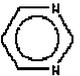 Pyrimidine | 11.4 [54] | 9.4 | 17.5 | -4.6 | 4.8 | 5.1 [56,104] | 6 |
| 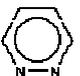 Pyridazine | 10.5 [54] | 9.4 | 10.5 | -4.6 | 4.8 | 4.9 [105]-5.0 [56,105] | 2-4 |
| 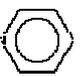 Pyridine | 9.7 [54]-9.8 [106] | 9.2 | 5.2-6.1 | -4.5 | 4.7 | 4.8 [56]-5.0 [104] | 2-6 |
| 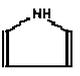 Pyrrole | 8.2 [107] | 8.3 | 1.2 | -3.8 | 4.5 | - | - |
| 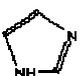 1H-imidazole | 8.8 [107]-9.0 [108] | 8.4 | 4.5-6.7 | -3.8 | 4.6 | - | - |
| 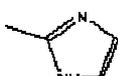 1H-imidazole,2methyl | 8.5 [108] | 8.4 | 1.2 | -3.9 | 4.5 | - | - |

*3.5 Organic compounds containing oxygen atoms*

Table V includes the theoretical results and the corresponding experimental data for cyclic and linear heteroatomic organic compounds containing oxygen atoms. For the non-cyclic molecules of this table the LCAO results for π orbitals can be readily obtained analytically because only two atoms are involved in π-bonding.

Regarding the πHOMO ionization energies of the depicted molecules the theoretical results do not differ more than 15% from the experimental values. The highest deviation is 14.5% in 2-pentanone, while for four molecules of Table V the deviation between the theoretical prediction and the experimental result does not exceed 10%. Similarly to the previous subsection 3.4, many molecules of Table V have an n orbital as HOMO. Specifically, this is known to be the case in acetone [109], acetaldehyde [110], 2-pentanone [111], and p-benzoquinone [112].

Looking at the π-π* energy gaps, larger deviations are obtained, except for the acetaldehyde where the deviation is around 9%. The relative errors in p-benzoquinone and 2,4-cyclopentadiene-1-one, which are the highest obtained in respect to all tables in this work, are about 49% and 52%, respectively. The main reason for which the LCAO method fails to predict the first π-π* transition in 2,4-cyclopentadiene-1-one, is the strong interaction of π orbitals with near degenerate states [36], which are ignored in the present treatment. Regarding p-benzoquinone two n states are found between the highest π and the lowest π* orbitals, and therefore the simple LCAO method used here is not able to give accurate results [113]. We mention at this point that the first π-π* transition is optically forbidden in this molecule [113]. For acetone, the first π-π* transition has not yet been clarified, due to the mixing of π orbitals with Rydberg or n states [114]. Information could not be found for the molecules of 2-pentanone and 2,4-cyclohexadien-1-one,6-methylene.

**Table V.** Organic compounds containing oxygen atoms. The columns represent the same quantities as in Table I.

| Organic molecule | $I\pi_{exp}$ (eV) | $I\pi_{th}$ (eV) | error$_1$ % | $L\pi_{th}$ (eV) | π-π*$_{th}$ (eV) | π-π*$_{exp}$ (eV) | error$_2$ % |
|---|---|---|---|---|---|---|---|
| 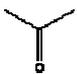 Acetone | 12.6 [109] | 13.4 | 6.3 | -5.1 | 8.3 | - | - |
| 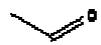 Acetaldehyde | 13.2 [110] | 13.4 | 1.5 | -5.1 | 8.3 | 9.1 [115] | 9 |
| 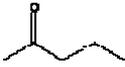 2-pentanone | 11.7 [111] | 13.4 | 14.5 | -5.1 | 8.3 | - | - |
| 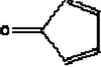 2,4-cyclopentadiene-1-one | 9.5 [116] | 8.5 | 10.5 | -6.9 | 1.6 | 3.3 [117] | 52 |
| 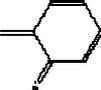 2,4-cyclohexadien-1-one, 6-methylene | 8.8 [118] | 8.9 | 1.1 | -6.3 | 2.6 | - | - |
| 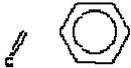 Benzaldehyde | 9.6 [58] -9.8 [119] | 9.2 | 4.2-6.1 | -5.7 | 3.5 | 4.4 [120] | 20 |
| 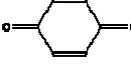 p-benzoquinone | 10.9 [112] | 9.4 | 13.8 | -7.3 | 2.1 | 4.1 [121] | 49 |

*3.6 Nucleic acids bases*

Molecules with crucial biological interest like the DNA and RNA bases are included in Table VI. Starting the discussion from the ionization energies of the highest occupied π orbital, we notice that good agreement with the experiment is obtained. The larger deviation is found in uracil (5.3-7.2%). In adenine and thymine the relative errors are smaller (2.4-3.5% and 0-2.2%, respectively), while in guanine and cytosine the present theoretical results are identical with the observed values.

Concerning the energy of the first π-π* transition, relatively small deviations are obtained. In adenine and guanine appear the larger relative errors, 16-17% and 12-17%, respectively. The lowest deviation is in cytosine, where the present LCAO prediction coincides with some of the experimental observations. Comparison between the experimental data and results from several first principles methods for all molecules of Table VI shows that the simple method used here is not much worse than the latter theoretical methods. In the case of adenine, the energy of the first π-π* transition, as predicted by several methods from first principles, is found in the range of 4.97-5.13 eV [122, 123, 124]. This overestimates the experimental value (4.5-4.6 eV) and it is only slightly better than the underestimated value predicted by our simple LCAO method (3.8 eV). Similar is the situation for guanine, where first principles methods evaluate the first π-π* energy in the region 4.76-4.96 eV [122, 125], overestimating the experimental result (4.3-4.6 eV), while the LCAO prediction is 3.8 eV. Regarding the molecule of cytosine the results from first principles methods are 4.39-4.71 eV [126, 127, 128], which are in very good agreement with the experiment (4.4-4.7 eV), as it is also the case for the accurate LCAO prediction (4.5 eV). Finally for the molecules of thymine and uracil methods from first principles predict the first π-π* transition in a region of 4.75-5.17 eV [125, 129, 130] and 4.82-5.44 eV [129, 130, 131, 132], respectively. The lower values in these regions are in better agreement with the experimental results, comparing to the LCAO method, but this is not the case for the higher predicted energies. It must be mentioned also that for these two molecules first principles calculations predict that the HOMO-LUMO transition is an n-π* transition. This does not seem to agree with the earlier established general acceptance that in all these five DNA and RNA bases the HOMO and LUMO are π orbitals [133].

**Table VI.** Nucleic acids bases. The columns represent the same quantities as in Table I.

| Organic molecule | $I\pi_{exp}$ (eV) | $I\pi_{th}$ (eV) | error$_1$ % | $L\pi_{th}$ (eV) | π-π*$_{th}$ (eV) | π-π*$_{exp}$ (eV) | error$_2$ % |
|---|---|---|---|---|---|---|---|
| 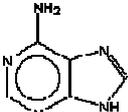 Adenine | 8.4 [133] -8.5 [134] | 8.2 | 2.4-3.5 | -4.4 | 3.8 | 4.5-4.6 [135-139] | 16-17 |
| 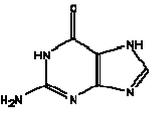 Guanine | 8.2 [133] | 8.2 | 0 | -4.4 | 3.8 | 4.3-4.6 [135,140-142] | 12-17 |
| 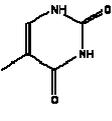 Thymine | 9.0- 9.2 [133,143,144] | 9.0 | 0-2.2 | -4.8 | 4.2 | 4.7-4.8 [137,145,146] | 11-13 |
| 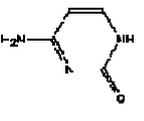 Cytosine | 8.9 [133] | 8.9 | 0 | -4.4 | 4.5 | 4.4-4.7 [135,137,139,141, 145,147-150] | 0-4 |
| 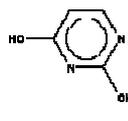 Uracil | 9.5-9.7 [133,143, 144,151] | 9.0 | 5.3-7.2 | -4.8 | 4.2 | 4.8 [137] | 13 |

## 4. Conclusions

Using the simplest form of the LCAO approximation, which takes into account only $p_z$ atomic orbitals, and a minimal set of unified parameters (four for the diagonal matrix elements corresponding to atoms of carbon ($\varepsilon_C = -6.7$ eV), nitrogen with coordination number 2 ($\varepsilon_{N_2} = -7.9$ eV), nitrogen with coordination number 3 ($\varepsilon_{N_3} = -10.9$ eV), and oxygen ($\varepsilon_O = -11.8$ eV), and the interatomic matrix elements between first neighboring $p_z$-type atomic orbitals proposed by Harrison [37]), the π molecular electronic structure of more than sixty planar organic molecules with sp$^2$ hybridized atoms has been evaluated. The energies of the πHOMO states and the πHOMO-πLUMO gaps have been compared with experimental data.

    The choice of the values of these four empirical parameters has been obtained through optimization in respect to the ionization energy of the highest occupied π molecular orbital. In particular, the value of $\varepsilon_C$ has been obtained first, by considering the molecules presented in Tables I-III (46 molecules totally). The resulting optimized value is almost the same as that obtained analytically considering the πHOMO of benzene. Then, keeping fixed this value of $\varepsilon_C$, the remaining values of $\varepsilon_{N_2}$, $\varepsilon_{N_3}$, and $\varepsilon_O$ have been optimized considering the molecules presented in Tables IV-VI (19 molecules totally), with special emphasis on the four DNA bases of Table VI. We mention that such an optimization is not a demanding computational process due to the simplicity of the method.

    Our theoretical calculations predict the experimental value of the πHOMO energy of sixty five organic compounds with a relative error of less than 15% in all cases examined, except for pyrimidine, where the deviation is 17.5%. Regarding the first π-π* transition, the deviations from the experimental observations are larger, but no more than about 50%, although there is not any adjustable parameter in this case (we reiterate that the optimization for obtaining the empirical parameters did not include the π-π* gaps). In forty five from the investigated molecules, experimental data for the first π-π* transition were available. In one case the deviation from the theoretical prediction is around 52%, in other five cases the deviation is between 40-50%, while in the other cases the relative error is below 40%.

Taking into account the simplicity of the method in respect to more accurate first principles calculations, the minimal number of parameters used, and the complicated actual electronic structure in some of the investigated cases (strong mixing between π, Rydberg and n states), it seems that the LCAO approach presented here provides a relatively accurate tool for a quick and easy derivation of theoretical estimates concerning the π electronic structure (at least the πHOMO and πLUMO states of interest) for planar organic molecules, which apart from carbon and hydrogen may also include nitrogen and/or oxygen atoms. This model computationally requires just a trivial diagonalization and can be easily used from no-specialists in hard theoretical or numerical calculations. We mention also that some cases can be treated even analytically within our approach. Compared to earlier Hückel approaches [38], our method offers improved predictions, at least for the molecular properties examined here. A restriction of the presented method is that it refers only to singlet electronic states, since only single-electron spatial wavefunctions are considered.

The rather accurate description of the HOMO and LUMO energies of DNA bases (the relative error is no more than 3.5% for HOMO orbitals and does not exceed 17% for the HOMO-LUMO transitions) suggests that the obtained wavefunctions can be used for estimating interbase coupling parameters (using appropriate atomic matrix elements [152] in a Slater-Koster type of coupling [153]), which are relevant for hole or electron transfer between DNA bases [154]. Such parameters can be used in phenomenological tight-binding descriptions of charge transfer along DNA. Work in this direction is in progress [155]. Another problem of biological interest that is related with the nature of π molecular orbitals and can be investigated within our approach, concerns the lowest excited state of flavin in the FADH$^-$ cofactor of the enzyme photolyase, which is involved in radiative DNA damage repair [156].

## References

[1] B. Alberts *et al*., "Essential Cell Biology", 2$^{nd}$ edition, Garland Science Publishing (2004).
[2] R. A. Morton and G. A. J. Pitt, Adv. Enzymol. Relat. Sub. & Biochem. **32**, 97 (1969); E. W. Abrahamson and S. E. Ostroy, Progr. Biophys. Mol. Biol. **17**, 179 (1967).
[3] G. Horowitz, Adv. Matter **10**, 365 (1998) and references therein.
[4] F. F. Heyroth and J. R. Loofbourow, J. Am. Chem. Soc. **56**, 1728 (1934).
[5] E. P. Carr and H. Stücklen, J. Chem. Phys. **4**, 760 (1936).
[6] V. Henri and L. W. Pickett, J. Chem. Phys. **7**, 439 (1939).
[7] J. R. Platt and H. B. Klevens, J.Chem. Phys. **16**, 832 (1948).
[8] J. R. Platt, H. B. Klevens, and W. C. Price, J. Chem. Phys. **17**, 466 (1949).
[9] H. B. Klevens and J. R. Platt , J. Chem. Phys. **17**, 470 (1949).
[10] R. A. Friedel, M. Orchin, and L. Reggel, J. Am. Chem. Soc. **70**, 199 (1948).
[11] L. Doub and J. M. Vandenbelt, J. Am. Chem. Soc. **71**, 2414 (1949).
[12] L. Doub and J. M. Vandenbelt, J. Am. Chem. Soc. **69**, 2714 (1947).
[13] F. Halverson and R. C. Hirt, J. Chem. Phys. **19**, 711 (1951); R. C. Hirt, F. Halverson, and R. G. Schmitt, J. Chem. Phys. **22**, 1148 (1954).
[14] J. T. Gary and L. W. Pickett, J. Chem. Phys. **22**, 599 (1954); L. C. Jones and L. W. Taylor, Analytical Chemistry **27**, 228 (1955); G. R. Hunt and I. G. Ross, J. Molecular Spectroscopy **9**, 50 (1962).
[15] P. A. Mullen and M. K. Orloff, J. Chem. Phys. **51**, 2277 (1969).
[16] W. M. Flicker, O. A. Mosher, and A. Kuppermann, Chem. Phys. Lett. **45**, 492 (1977).
[17] C. D. Cooper *et al*., J. Chem. Phys. **73**, 1527 (1980).

[18] I. C. Walker, Chem. Phys. **109**, 269 (1986).
[19] M. H. Palmer *et al.*, Chem. Phys. **117**, 51 (1987).
[20] A. G. Robinson *et al.*, J. Chem. Phys. **116**, 7918 (2002).
[21] E. Hückel, Z. Phys. **70**, 204 (1931); Z. Phys. **76**, 628 (1932).
[22] W. Kutzelnigg, J. Comput. Chem. **28**, 25 (2007).
[23] J. R. Platt, J. Chem. Phys. **18**, 1168 (1950).
[24] J. R. Platt, J. Chem. Phys. **19**, 101 (1951).
[25] R. Pariser, and R. G. Parr, J. Chem. Phys. **21**, 466 (1953).
[26] R. Pariser, and R. G. Parr, J. Chem. Phys. **21**, 767 (1953).
[27] R. Pariser, J. Chem. Phys. **24**, 250 (1956).
[28] N.O. Lipari and C. B. Duke, J. Chem. Phys. **63**, 1768 (1975).
[29] J. E. Del Bene, J. D. Watts, and R. J. Bartlett, J. Chem. Phys. **106**, 6051 (1997).
[30] H. H. Heinze, A. Görling, and N. Rösch, J. Chem. Phys. **113**, 2088 (2000).
[31] E. S. Kadantsev, M. J. Stott, and A. Rubio, J. Chem. Phys. **124**, 134901 (2006).
[32] R. G. Endres, C. Y. Fong, L. H. Yang, G. Witte, and Ch. Wöll, Computational Materials Science **29**, 362 (2004).
[33] A. B. Trofimov and J. Schirmer, Chem. Phys. **214**, 153 (1997); O. Christiansen *et al.*, J. Chem. Phys. **111**, 525 (1999).
[34] J. Wan *et al.*, J. Chem. Phys. **113**, 7853 (2000).
[35] M. Boggio-Pasqua *et al.*, J. Chem. Phys. **120**, 7849 (2004).
[36] L. Serrano-Andres *et al.*, J. Chem. Phys. **117**, 1649 (2002).
[37] W. A Harrison, "Electronic structure and the properties of solids'', 2[nd] edition, Dover, New York (1989); W. A Harrison, "Elementary electronic structure'', World Scientific (1999).
[38] A. Streitwieser, "Molecular Orbital Theory for Organic Chemists'', Wiley (1961).
[39] T. Cramer, S. Krapf, and T. Koslowski, J. Phys. Chem. B **108**, 11812 (2004); T. Cramer, S. Krapf, and T. Koslowski, J. Phys. Chem. C **111**, 8105 (2007).
[40] M. Wolfsberg and L. Helmholz, J. Chem. Phys. **20**, 837 (1952).
[41] NIST Chemistry Webbook (http://webbook.nist.gov/chemistry/) and references therein.
[42] B. O. Roos, M. P. Fulscher, P.-A. Malmqvist, M. Merchan, and L. Serrano-Andres, "Quantum Mechanical Electronic Structure Calculations with Chemical Accuracy'', edited by S. R. Langhoff, Kluwer Academic, Amsterdam (1995).
[43] S. Grimme and M. Parac, Tech. Rep. Ser. - I. A. E. A. **4**, 292 (2003).
[44] J. O. Howell, J. M. Goncalves *et al.*, J. Am. Chem. Soc. **106**, 3968 (1984).
[45] B. Kovac, M. Mohraz *et al.*, J. Am. Chem. Soc., **102**, 4314 (1980).
[46] W. Kaim, H. Tesmann, H. Bock, Chem. Ber. **113**, 3221 (1980).
[47] T. Kobayashi, Phys. Lett. **69**, 105 (1978).
[48] L. Klasinc, I. Novak, M. Scholz, and G. Kluge, Croat. Chem. Acta. **51**, 43 (1978).
[49] W. Schmidt, J. Chem. Phys. **66**, 828 (1977).
[50] E. Clar and W. Schmidt, Tetrahedron. **32**, 2563 (1976).
[51] P. K. Bischof, M. J. S. Dewar, D. W. Goodman, and T. B. Jones, J. Organomet. Chem. **82**, 89 (1974).
[52] M. Klessinger, Angew. Chem. Int. Ed. Engl. **11**, 525 (1972).
[53] H. Bock, G. Wagner, J. Kroner, Chem. Ber. **105**, 3850 (1972).
[54] R. Gleiter, E. Heilbronner, and V. Hornung, Angew. Chem. Int. Edit. **9**, 901 (1970).
[55] J. A. Sell and A. Kupperman, Chem. Phys. **33**, 367 (1978); M. Gower, L. A. P. Kane-Maguire, J. P. Maier, and D. A. Sweigart, J. Chem. Soc. Dalton Trans. 316. (1977); H. Bock, W. Kaim, and H. E. Rohwer, J. Organomet. Chem. **135**, 14 (1977); T. Kobayashi and S. Nagakura, J. Electron Spectrosc. Relat. Phenom. **7**, 187 (1975);

W. Schafer, A. Schweig, Angew. Chem. **84**, 898 (1972); T. A. Carlson and C. P. Anderson, Chem. Phys. Lett. **10**, 561 (1971); H. Bock, W. Fuss, Angew. Chem. Int. Ed. Engl. **10**, 182 (1971).
[56] C. N. R. Rao, "Ultra-violet and visible spectroscopy: chemical applications'', Butterworth (1975).
[57] H. H. Perkampus, "UV-VIS Atlas of Organic Compounds", VCH, Weinheim (1992).
[58] L. Klasinc, B. Kovac, and H. Gusten, Pure Appl. Chem. **55**, 289 (1983).
[59] E. Heilbronner, T. Hoshi, J. L. von Rosenberg, and K. T. Hafner, Nouv. J. Chim. **1**, 105 (1976).
[60] F. Marschner and H. Goetz, Tetrahedron. **30**, 3159 (1974).
[61] H. Bock and G. Wagner, Angew. Chem. Int. Ed. Engl. **11**, 119 (1972).
[62] N. S. Hush, A. S. Cheung, and P. R. Hilton, J. Electron Spectrosc. Relat. Phenom. **7**, 385 (1975)
[63] L. Klasinc, B. Kovac, S. Schoof, and H. Gusten, Croat. Chem. Acta. **51**, 307 (1978); C. Jongsma, *et al*, Tetrahedron. **31**, 2931 (1975); W. Schafer, A. Schweig, F. Bickelhaupt, and H. Vermeer, Angew. Chem. Int. Ed. Engl. **11**, 924 (1972).
[64] I. Akiyama, K. C. Li *et al*, J. Phys. Chem. **83**, 2997 (1979).
[65] B. Ruscic, B. Kovac, L. Klasinc, and H. Gusten, Z. Naturforsch. A:. **33**, 1006 (1978).
[66] J. P. Maier and D. W. Turner, Faraday Discuss. Chem. Soc. **54**, 149 (1972).
[67] F. Brogli and E. Heilbronner, Angew. Chem. Int. Ed. Engl. **11**, 538 (1972).
[68] D. Biermann and W. Schmidt, J. Am. Chem. Soc. **102**, 3173 (1980).
[69] N. Nijegorodov, V. Ramachandran, and D. P. Winkoun, Spectrochimica Acta Part A. **53**, 1813 (1997).
[70] E. Clar and W. Schmidt, Tetrahedron. **33**, 2093 (1977); R. Boschi and W. Schmidt, Tetrahedron Lett. **25**, 2577 (1972).
[71] C. Pentrongolo, R. J. Buenker, and S. D. Peyerimhoff, J. Chem. Phys. **76**, 3655 (1982).
[72] L. E. McMurchie and E. R. Davidson, J. Chem. Phys. **67**, 5613 (1977).
[73] R. Lindh and B. O. Roos, Int. J. Quantum Chem. **35**, 813 (1989).
[74] L. Serrano- Andres et al., J. Chem. Phys. **98**, 3151 (1993).
[75] A. Kuppermann, W. M. Flicker, and O. A. Mosher, Chem. Reviews. **79**, 77 (1979).
[76] D. A. Krause, J. W. Taylor, and R. F. Fenske, J. Am. Chem. Soc. **100**, 718 (1978).
[77] R. M. White, T. A. Carlson, and D. P. Spears, J. Electron Spectrosc. Relat. Phenom. **3**, 59 (1974).
[78] G. Bieri and L. Asbrink, J. Electron Spectrosc. Relat. Phenom. **20**, 149 (1980).
[79] J. Merer and R. S. Mulliken, Chem. Reviews **69**, 639 (1969).
[80] K. Kimura, S. Katsumata, T. Yamazaki, and H. Wakabayashi, J. Electron Spectrosc. Relat. Phenom. **6**, 41 (1975).
[81] G. Hentrich, E. Gunkel, and M. Klessinger, J. Mol. Struct. **21**, 231 (1974); U. Weidner and A. Schweig, J. Organomet. Chem. **39**, 261 (1972); P. Mollere, H. Bock, G. Becker, and G. Fritz, J. Organomet. Chem. **46**, 89 (1972).
[82] I. C. Walker *et al.*, Chem. Phys. **109**, 269 (1986).
[83] K. B. Wiberg, G. B. Ellison *et al*, J. Am. Chem. Soc. **98**, 7179 (1976); T. Koenig, T. Balle, and W. Snell, J. Am. Chem. Soc. **97**, 662 (1975).
[84] M. H. Palmer *et al.*, Chem. Phys. **117**, 51 (1987).
[85] E. J. McAlduff and K. N. Houk, Can. J. Chem. **55**, 318 (1977); P. D. Mollere, K. N. Houk, D. S. Bomse, and T. H. Morton, J. Am. Chem. Soc. **98**, 4732 (1976); W. Fuss and H. Bock, J. Chem. Phys. **61**, 1613 (1974).
[86] H. Schmidt and A. Schweig, Tetrahedron. **32**, 2239 (1976).

[87] O. A. Mosher, W. M. Flicker, and A. Kuppermann, Chem. Phys. Lett. **19**, 332 (1973); J. Chem. Phys. **59**, 6502 (1973).
[88] R. McDiarmid, Chem. Phys. Lett. **34**, 130 (1975).
[89] R. M. Gavin and S. A. Rice, J. Chem. Phys. **60**, 3231 (1974).
[90] F. Marschner and H. Goetz, Tetrahedron. **30**, 3451 (1974); T. Kobayashi and S. Nagakura, Chem. Lett. 903 (1972).
[91] J. B. Lambert, H. F. Shurvell, and D. A. Liqhtner, "Organic Stuctural Spectroscopy'' Prentice-Hall (1998).
[92] P. Bruckmann and M. Klessinger, Chem. Ber. **107**, 1108 (1974); T. Kobayashi, K. Yokota, and S. Nagakura, J. Electron Spectrosc. Relat. Phenom. **3**, 449 (1973); E. W. Fu and R. C. Dunbar, J. Am. Chem. Soc. **100**, 2283 (1978).
[93] J. Wan and H. Nakatsuji, Chem. Phys. **302**, 125 (2004).
[94] T. Koenig, M. Tuttle, and R. A. Wielesek, Tetrahedron Lett. 2537 (1974).
[95] F. Brogli, E. Giovannini, E. Heilbronner, and R. Schurter, Chem. Ber. **106**, 961 (1973).
[96] R. Gleiter, H. Hopf, M. Eckert-Maksic, and K. L. Noble, Chem. Ber. **113**, 3401 (1980).
[97] T. Koenig, R. Wielesek, W. Snell, and W. Balle, J. Am. Chem. Soc. **97**, 3225 (1975).
[98] J. P. Maier and D. W. Turner, J. Chem. Soc. Faraday Trans. 2, **69**, 196 (1973).
[99] R. Gleiter, W. Schafer, and M. Eckert-Maksic, Chem. Ber. **114**, 2309 (1981) ; Y. V. Chizhov, M. M. Timoshenko *et al*, J. Struct. Chem. **27**, 401 (1986).
[100] D. Dougherty, J. Lewis, R. V. Nauman, and S. P. McGlynn, J. Electron Spectrosc. Relat. Phenom. **19**, 21 (1980).
[101] Y. Lou *et al*., J. Am. Chem. Soc. **124**, 15302 (2002).
[102] T. Kobayashi, Bull. Chem. Soc. Jpn. **56**, 3224 (1983).
[103] M. Shahbaz *et al*., J. Am. Chem. Soc. **106**, 2805 (1984).
[104] A. Bolovinos, P. Tsekeris, J. Philis, E. Pantos, and G. Andritsopoulos, J. Mol. Spectroscopy **103**, 240 (1984).
[105] K. K. Innes, I. G. Ross, and W. R. Moomaw, J. Mol. Spectroscopy **132**, 492 (1988).
[106] C. Batich, E. Heilbronner *et al*, J. Am. Chem. Soc. **95**, 928 (1973); E. Heilbronner, V. Hornung, F. H. Pinkerton, and S. F. Thames, Helv. Chim. Acta. **55**, 289 (1972).
[107] S. Cradock, R. H. Findlay, and M. H. Palmer, Tetrahedron. **29**, 2173 (1973).
[108] B. G. Ramsey, J. Org. Chem. **44**, 2093 (1979).
[109] V. Y. Young and L. Cheng, J. Chem. Phys. **65**, 3187 (1976).
[110] K. Johnson, I. Powis, and C. J. Danby, Chem. Phys. **70**, 329 (1982).
[111] P. R. Olivato *et al.*, J. Chem. Soc. Perkin Trans. II, 1505 (1984).
[112] D. Dougherty and S. P. McGlynn, J. Am. Chem. Soc. **99**, 3234 (1977).
[113] R. P. Amerigo, M. Merchán, and E. Orti, J. Chem. Phys. **110**, 9536 (1999).
[114] M. Merchán *et al*., J. Chem. Phys. **104**, 1791 (1996).
[115] C.R. Silva and J.P. Reilly, J. Phys. Chem. **100**, 17111 (1996).
[116] T. Koenig, M. Smith, and W. Snell, J. Am. Chem. Soc. **99**, 6663 (1977).
[117] E. W. Garbisch Jr. and R. F. Sprecher, J. Am. Chem. Soc. **88**, 3433 (1966).
[118] V. Eck, A. Schweig, *et al*., Tetrahedron Lett. **27**, 2433 (1978).
[119] A. D. Baker, D. P. May, and D.W. Turner, J. Chem. Soc. B 22, (1968)
[120] C. M. Hadal, J. B. Foresman, and K. B. Wiberg, J. Phys. Chem. **97**, 4293 (1993).
[121] H. P. Trommsdorff, J. Chem. Phys. **56**, 5358 (1972).
[122] M. P. Fűlscher, L. Serrano-Andrés, and B. O. Roos, J. Am. Chem. Soc. **119**, 6168 (1997).

[123] B. Mennucci, A. Toniolo, and J. Tomasi, J. Phys. Chem. A. **105**, 4749 (2001).
[124] A. L. Sobolewski and W. Domcke, Eur. Phys. J. D **20**, 369 (2002).
[125] C. E. Crespo-Hernández, C. Marai, and B. Kohler, unpublished results.
[126] N. Ismail, L. Blancafort, M. Olivucci, B. Kohler, and M. A. Robb, J. Am. Chem. Soc. **124**, 6818 (2002).
[127] M. K. Shukla and J. Leszczynski, J. Phys. Chem. A. **106**, 11338 (2002).
[128] M. P. Fűlscher and B. O. Roos, J. Am. Chem. Soc. **117**, 2089 (1995).
[129] J. Lorentzon, M. P. Fűlscher, and B. O. Roos, J. Am. Chem. Soc. **117**, 9265 (1995).
[130] M. K. Shukla and P. C. Mishra, Chem. Phys. **240**, 319 (1999).
[131] A. Broo, G. Pearl, and M. C. Zerner, J. Phys. Chem. A **101**, 2478 (1997).
[132] C. M. Marian, F. Schneider, M. Kleinschmidt, and J. Tatchen, Eur. Phys. J. D **20**, 357 (2002).
[133] N. S. Hush and A. S. Cheung, Chem. Phys. Lett. **34**, 11 (1975).
[134] J. Lin, C. Yu *et al*, J. Am. Chem. Soc. **102**, 4627 (1980) ; S. Peng, A. Padva, and P. R. LeBreton, Proc. Nat. Acad. Sci. U.S.A. **73**, 2966 (1976).
[135] D. Voet, W. B. Gratzer, R. A. Cox, and P. Doty, Biopolymers **1**, 193 (1963).
[136].T. Yamada and H. Fukutome, Biopolymers **6**, 43 (1968).
[137] Y. Matsuoka and B. Norden, J. Phys. Chem. **86**, 1378 (1982).
[138] L. B. Clark, J. Phys. Chem. **94**, 2873 (1990).
[139] W. Voelter, R. Records, E. Bunnenberg, and C. Djerassi, J. Am. Chem. Soc. **90**, 6163 (1968).
[140] L. B. Clark, J. Am. Chem. Soc. **99**, 3934 (1977).
[141] C. A. Sprecher and W. C. Johnson, Biopolymers **16**, 2243 (1977).
[142] J. C. Sutherland and K. Griffin, Biopolymers **23**, 2715 (1984).
[143] D. Dougherty, K. Wittel, J. Meeks, and S. P. McGlynn, J. Am. Chem. Soc. **98**, 3815 (1976).
[144] G. Lauer, W. Schafer, and A. Schweig, Tetrahedron Lett. **45**, 3939 (1975).
[145] L. B. Clark and I. Tinoco Jr., J. Am. Chem. Soc. **87**, 11 (1965).
[146] L. B. Clark, G. G. Peschel, and I. Tinoco Jr., J. Phys. Chem. **69**, 3615 (1965).
[147] K. Raksanyi, I. Foldvary, J. Fidy, and L. Kittler, Biopolymers **17**, 887 (1978).
[148] T. P. Lewis and W. A. Eaton, J. Am. Chem. Soc. **93**, 2054 (1971).
[149] F. Zaloudek, J. S. Novros, and L. B. Clark, J. Am. Chem. Soc. **107**, 7344 (1985).
[150] A. F. Fucaloro and L. S. Forster, J. Am. Chem. Soc. **93**, 6443 (1971).
[151] M. H. Palmer, I. Simpson, and R. J. Platenkamp, J. Mol. Struct. **66**, 243 (1980); A. Padva, P. R. LeBreton, R. J. Dinerstein, and J. N. A. Ridyard, Biochem. Biophys. Res. Commun. **60**, 1262 (1974).
[152] M. Menon and R. E. Allen, Phys. Rev. B **38**, 6196 (1988); N. Lathiotakis and A. N. Andriotis, Solid State Comm. **87**, 871 (1993); M. Menon, J. Connolly, N. Lathiotakis, and A. Andriotis, Phys. Rev. B **50**, 8903 (1994).
[153] J. C. Slater and G. F. Koster, Phys. Rev. **94**, 1498 (1954).
[154] R. G. Endres, D. L. Cox, and R. R. P. Singh, Rev. Mod. Phys. **76**, 195 (2004).
[155] L. Hawke, G. Kalosakas, and C. Simserides, in preparation.
[156] T. R. Prytkova, D. N. Beratan, and S. S. Skourtis, Proc. Natl. Acad. Sci. **104**, 802 (2007).